\documentclass[%
 reprint,
 amsmath,amssymb,
 aps,
]{revtex4-1}

\usepackage{graphicx}
\usepackage{dcolumn}
\usepackage{bm}
\usepackage{siunitx}
\usepackage{textcomp}
\usepackage{xcolor}

\begin{document}

\title{Information transfer by quantum matterwave modulation}

\author{R. R\"{o}pke$^1$, N. Kerker$^1$ and A. Stibor$^{1,2*}$}
\affiliation{\mbox{$^1$Institute of Physics and LISA$^+$, University of T\"{u}bingen, T\"{u}bingen, Germany} \mbox{$^2$Lawrence Berkeley National Lab, Molecular Foundry, Berkeley, USA, *email: astibor@lbl.gov}}

\begin{abstract}
\noindent Classical communication schemes that exploit wave modulation are the basis of the information era. The transfer of information based on the quantum properties of photons revolutionized these modern communication techniques. Here we demonstrate that also matterwaves can be applied for information transfer and that their quantum nature provides a high level of security. Our technique allows transmitting a message by a non-trivial modulation of an electron matterwave in a biprism interferometer. The data is encoded by a Wien filter introducing a longitudinal shift between separated matterwave packets. The transmission receiver is a delay line detector performing a dynamic contrast analysis of the fringe pattern. Our method relies on the Aharonov-Bohm effect and has no light optical analog since it does not shift the phase of the electron interference. A passive eavesdropping attack will cause decoherence and terminating the data transfer. This is demonstrated by introducing a semiconducting surface that disturbs the quantum state by Coulomb interaction and reduces the contrast. We also present a key distribution protocol based on the quantum nature of the matterwaves that can reveal active eavesdropping.
\end{abstract}

\maketitle
\setlength{\parindent}{0cm}
{\bf \large INTRODUCTION} 

Major foundations of quantum physics are uncertainty, randomness and entanglement. In the last decades, fascinating quantum information science schemes were developed based on those properties using photons, atoms, molecules or ions \cite{Bennett2000,Bennett1984,Beth2006}. Exploiting this quantum behavior, generated e.g.~in a parametric down conversion source \cite{Kwiat1995}, led to major achievements in quantum optics, including the violation of Bell inequality \cite{Ou1988}, quantum teleportation \cite{Bouwmeester1997} and quantum computing \cite{Walther2005} or secure data transmission such as quantum cryptography \cite{Jennewein2000}. 

\setlength{\parindent}{5mm}

An other quantum cornerstone, the wave-particle duality, generated various matterwave experiments and applications for electrons \cite{Hasselbach2010}, ions \cite{Hasselbach1999}, atoms \cite{cronin2009}, molecules \cite{Haslinger2013} and neutrons \cite{Rauch2015}. It was shown that almost any classical wave phenomenon has its counterpart realization with matterwaves leading to particle interference \cite{gerlich2007} and Bragg scattering \cite{Freimund2002} of matter on light gratings, atomic Mach-Zehnder interferometer \cite{Berrada2013}, transmission of electrons on bulk structures for tomography \cite{Xu2015} or even an atom laser by releasing atoms from a Bose-Einstein condensate \cite{Bloch1999}. However, matterwaves have yet not been exploited as carrier waves for the modulation and transmission of a signal, as it is common for classic electromagnetic data transfer. Extending such methods to quantum matterwaves has the potential of a significant improvement in security and may opens the door for a new type of communication scheme. The realization of non-local quantum information transducing architectures based on matterwaves was yet not possible due to a lack of a suitable transmitter element. 

\begin{figure*}[t]
\centering
\includegraphics[width=0.8\textwidth]{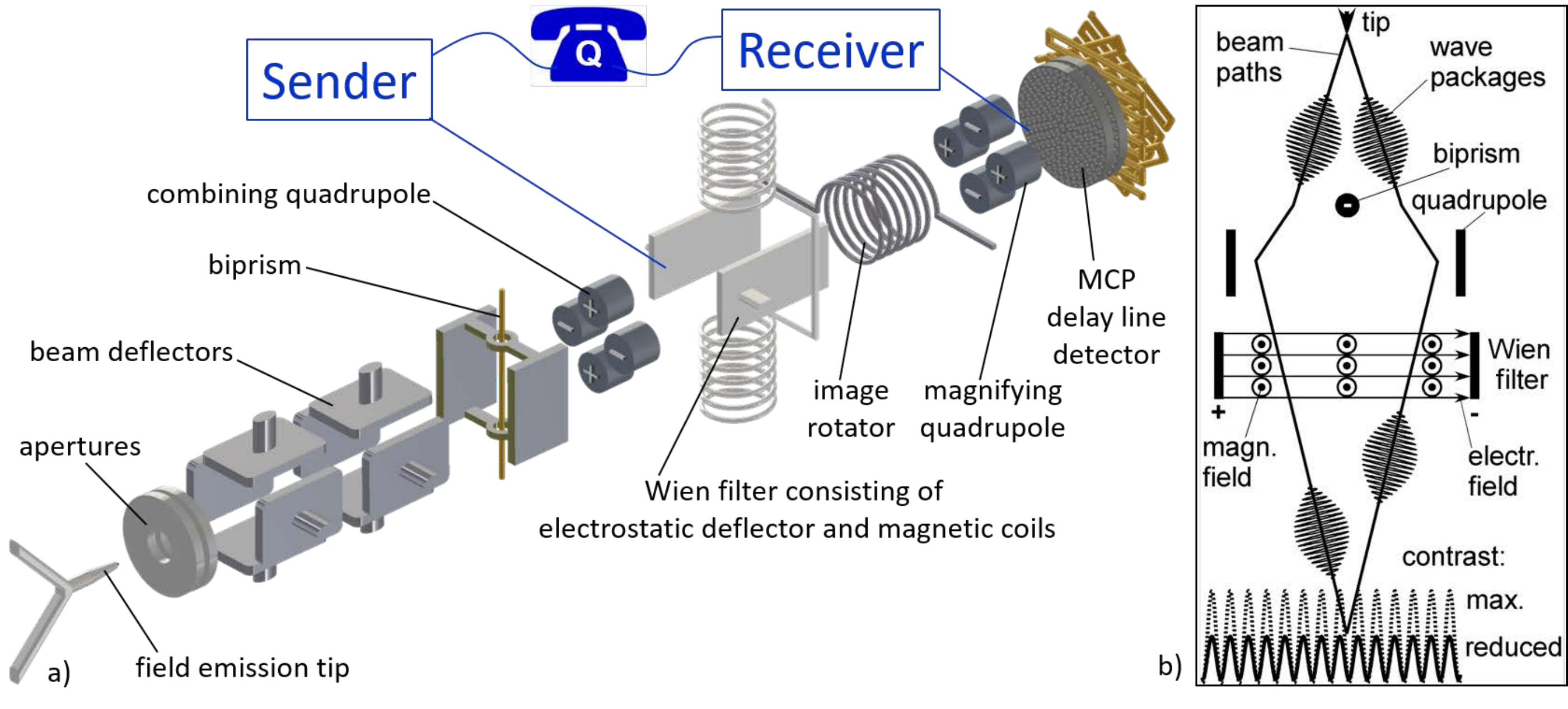} \caption{a) Experimental setup for the quantum communication in an electron interferometer. The sender encodes the message on the matterwave with a Wien filter, the receiver decodes it on the detector. A possible eavesdropper intercepting between the Wien filter and the magnifying quadrupole would cause Coulomb decoherence \cite{Sonnentag2007} that shuts down the conversation. b)~Sketch of the Wien filter in the matched mode encoding a binary signal on the spatially separated matterwave packets by longitudinal shifts between large and small overlap. The corresponding high and low interference fringe contrast is read out dynamically in a delay line detector. 
}
\label{figure1}
\end{figure*}

In this paper we demonstrate such an approach, where a message is transmitted by a non-trivial quantum modulation of an electronic matterwave. It cannot be realized with classical particles or light. We introduce and experimentally realize a unique information transfer scheme that is fundamentally different to the current quantum information science techniques with photons \cite{Yuan2010}. The matterwaves of electrons in a biprism interferometer \cite{Mollenstedt1956,Hasselbach2010,Rembold2014,Schuetz2014,Guenther2015,Pooch2017,Rembold2017} are used as the carrier waves to be modulated for signal transmission. The coherent electron beam is generated by a nanotip field emitter \cite{Kuo2006,Kuo2008,Pooch2018} and separated by a biprism fiber into two partial waves. After superposition by a quadrupole lens, an interference fringe pattern is formed on a delay line detector with a high spatial and temporal resolution \cite{Jagutzki2002}. Due to the limited energy width of the source, each detected electron has a slightly different wavelength. The ensemble of these matterwaves can be described by a wave packet with a distinct width being the longitudinal coherence length \cite{Nicklaus1993,Sonnentag2000}. With a Wien filter in the beam path the separated wave packets can be longitudinally shifted towards each other decreasing the interference contrast \cite{Hasselbach2010,Nicklaus1993,Sonnentag2000,Sonnentag2005} without changing the classical electron pathways or the phase of the fringe pattern. The Wien filter modulation is based on the Aharonov-Bohm effect on charged particle waves and therefore has no counterpart with photons or neutral atoms. The signal to be transmitted is encoded binary on the matterwave with the Wien filter by switching between high and low contrast. In this way, neither the beam position or total intensity, nor the interference phase are changed revealing it as a truly quantum modulation dependent on the electrons wave nature. The data readout is realized by a dynamic contrast measurement. A transfer rate of \SI{1}{bit} per \SI{5}{s} was achieved with a transmission distance in the centimeter range until overlap of the separated beams. The total transmission length to the detector was \SI{14}{cm}. The signal transmission is demonstrated to be secure for decohering interceptions by introducing a semiconducting plate parallel to the electron beam paths. The Coulomb interaction between beam electrons and image charges in the semiconductor represents an passive eavesdropping attempt. The reciprocal action causes a transfer of which-path information which quenches the interference fringes due to decoherence \cite{Sonnentag2007}, effectively annihilating the communication. We furthermore present a key distribution scheme for matterwaves with similarities to the BB84 protocol for photons \cite{Bennett1984}. It can reveal an active eavesdropping attack based on the quantum wave nature of the electrons. \\

\setlength{\parindent}{0cm}

{\bf \large RESULTS} 

The principles of electron biprism interferometry are explained in \cite{Mollenstedt1956,Hasselbach2010}. Our setup and its application for communication by matterwave modulation is illustrated in Fig.~\ref{figure1} a) and b). It includes several beam optic components from a former experiment by Sonnentag et al.~\cite{Sonnentag2007}. A single atom nanotip field emitter  \cite{Kuo2006,Kuo2008,Schuetz2014,Pooch2018} is the origin of coherent electron waves. The beam is guided by two electrostatic deflectors to illuminate the biprism. It consists of a Au/Pd-coated glass fiber with a diameter of \SI{395}{nm} \cite{Schuetz2014} and is placed between two grounded parallel plates. The fiber is set on a low negative potential between 0 and \SI{-1}{V} and can separate the beam coherently up to $\Delta x = $ \SI{8.5}{\micro\m}. The partial waves are combined again by a quadrupole lens \cite{Sonnentag2007,Schuetz2015b}. Before they are superposed at the entrance of a quadrupole magnification lens, they pass a Wien filter \cite{Nicklaus1993} where the signal to be transmitted gets modulated on the electron waves as described below. The beam additionally traverses a magnetic coil as an image rotator to adjust the edges of the overlapping partial beams normal to the magnifying direction of the quadrupole. After superposition, the resulting fringe pattern is magnified by a factor of about 1500 and detected on a delay line detector with a high spatial and temporal single particle resolution \cite{Jagutzki2002}. 
\setlength{\parindent}{5mm}

The non-zero energy width of the beam source $\Delta E$ and the quantum superposition principle allow to describe the beam with wave packets being the Fourier sum of individual linearly independent plane de Broglie waves of single electrons with slightly different energies \cite{Hasselbach2010,Nicklaus1993}.

The Wien filter is used in classical physics as a velocity filter and energy analyzer for electron beams \cite{Wien1897,Boersch1964}. In biprism electron interferometry the unit was rediscovered to correct for longitudinal wave packet shifts \cite{Nicklaus1993}. They are introduced by electrostatic deflectors for beam alignment. The partial waves traverse in regions between the deflector plates with different electrostatic potentials. It leads to non-equal group velocities and different arrival times of the wave packets on the detector. This in turn results in a longitudinally shifted overlap and a loss of contrast \cite{Hasselbach2010,Nicklaus1993,Sonnentag2000,Sonnentag2005}. When the shift is greater than the width of the wave packets, defined as the longitudinal coherence length, then the fringe contrast vanishes. The Wien filter can compensate for this relative delay and reestablish maximum contrast \cite{Nicklaus1993}. It consists of an electrostatic deflector and two connected magnetic coils generating a magnetic and electric field both perpendicular to each other and the direction of the beam path (see Fig.~\ref{figure1}~a)). The Wien filter is always used in a matched state where the effects of the electric and magnetic forces on the electrons cancel each other. In this so called matched mode, the Wien filter does not deflect the beam or shift the phase of the interference pattern, it has therefore no classical influence on the electrons. However, the spatially separated partial wave packets still experience different potentials between the Wien filter deflector plates which change their relative group velocities. The magnetic field does not introduce such an effect. For that reason, a combination of the electric and magnetic field can always be found to fully compensate the original longitudinal shift of the other deflectors in the beam line and to restore maximum contrast \cite{Nicklaus1993} without beam deflection. 

\begin{figure}[t]
\centering
\includegraphics[width=0.5\textwidth]{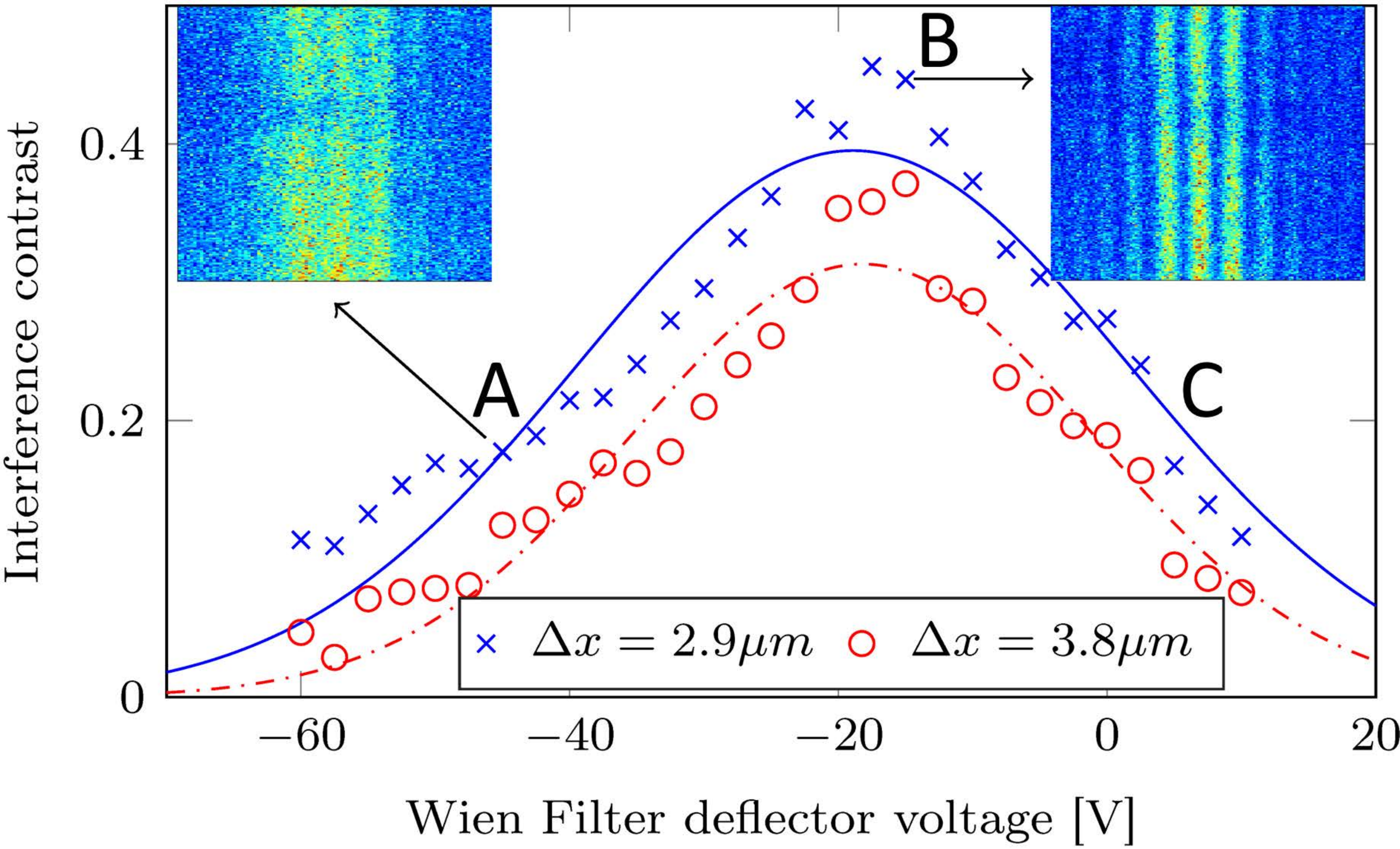} \caption{Dependency of the interference contrast with the Wien filter's deflector voltage in the matched mode for two different beam path separations, $\Delta x = $ \SI[separate-uncertainty = true]{2.9 \pm 0.4}{\micro\metre} (blue crosses) and $\Delta x = $ \SI[separate-uncertainty = true]{3.8 \pm 0.4}{\micro\metre} (red circles). The data is fitted by Gaussian distributions (solid blue and chain dotted red line). Insets: The determined electron fringe pattern on the detector for low contrast state 1 (left, point A) and high contrast state 2 (right, point B) in the signal transmission process.}
\label{figure2}
\end{figure}

In this paper, we use this mechanism to switch between two states in the Wien filter for binary signal transmission. The state 2 provides a full longitudinal overlap of the separated wave packets on the detector and maximal contrast. For state 1, a voltage on the Wien filter deflector is applied (and the according current through the magnetic coils for a matched state) to longitudinally shift the wave packets towards each other slightly less than the coherence length. This results in a significantly reduced interference contrast. The two states and the dependency of the measured electron interference contrast as a function of the matched Wien filter deflector voltage is shown in Fig.~\ref{figure2} for two different beam path separations. The two insets indicate the interference pattern visible on the detector at those points (labeled as A and B in Fig.~\ref{figure2}) which are switched for the binary encoding in the signal transmission. They reveal also the constant phase position of the fringe pattern.

\begin{figure*}[t]
\centering
\includegraphics[width=1.0\textwidth]{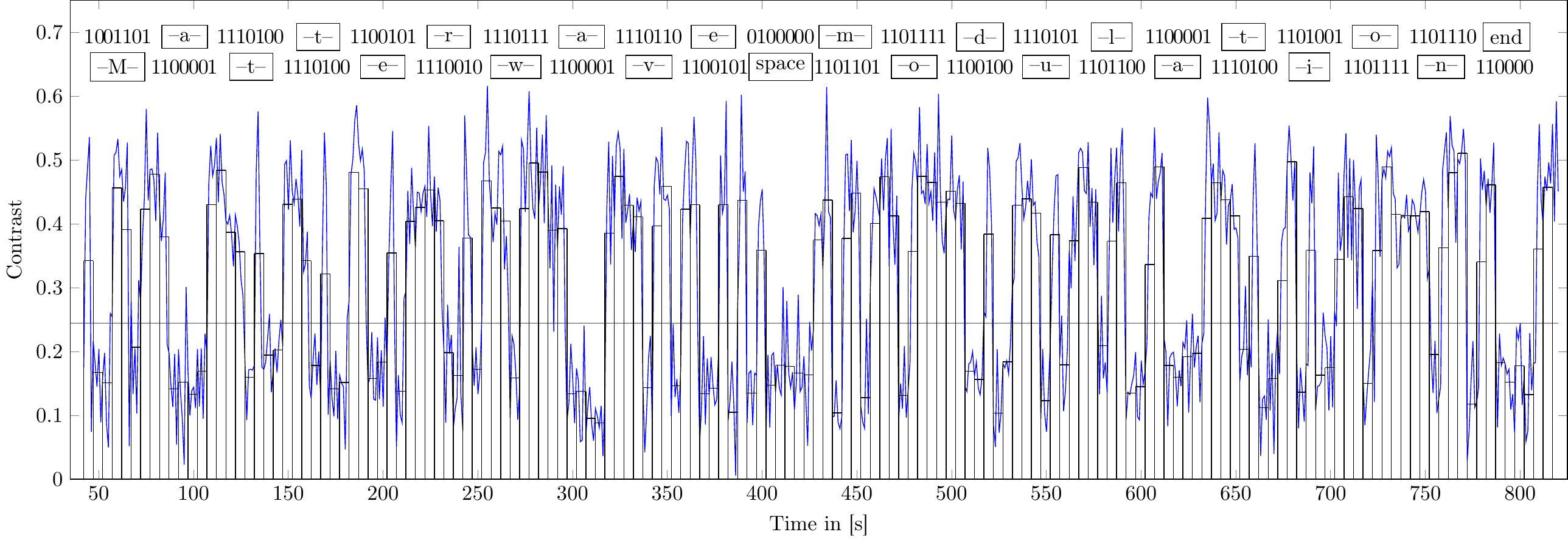} \caption{Interference contrast sequence for the transmitted message "Matterwave modulation". The blue curve indicates the contrast value for every single interferogram per time bin. The black rectangular bars are the average of five time bins, defining the contrast of a single bit. The red line is the cut-off for the contrast to be interpreted as a "0" or a "1" by the readout software. It also translates the bins from binary to letters, revealing the correctly transmitted message on top of the diagram. }
\label{figure3}
\end{figure*}

With this setup the first proof-of-principle experiment for information transfer by matterwave modulation was performed. The results are presented in Fig.~\ref{figure3}. A user interface was coded where we input the message "Matterwave modulation". It was transmitted from the sender (Wien filter) to the receiver (detector) as indicated in Fig.~\ref{figure1}. The program converts every letter in its binary representation and sends it to the control unit for the Wien filter. It consists of a pair of bipolar voltage sources for the Wien deflector, a bipolar current source for the Wien coils and a micro controller. For the binary number "1", the deflector voltage and coil current were set to state 2 (point B), for the binary number "0" they were set to state 1 (point A), as indicated in Fig.~\ref{figure2}. At these two points, the corresponding high or low contrast fringe pattern were recorded. Each interferogram reveals a particle event integration over one second. This time binning is a balance between the count rate and the signal transmission rate. To avoid errors, enough counts are needed to make sure the contrast and phase can be determined. Thus, 4000 to 5000 counts/s were accumulated on the whole screen, resulting in 1000 to 1500 counts/s for each recorded interference pattern within five fringes. It is sufficient signal to determine the fit parameters and the contrast with a reasonable accuracy as described in the methods \cite{Pooch2017,Pooch2018}. 
The extracted interference contrast is plotted by the blue curve in Fig.~\ref{figure3}. Five time bins are averaged to form one bit and the averaged values are represented by black rectangular bars. The message could thus be transmitted with a rate of 1 bit per 5 sec. To determine if a bit is "0" or "1", a cut-off value was defined as (red line): $C_{cutoff}=0.8\times\bar{C}$, with $\bar{C}$ being the average contrast of all interferograms in the transmission. The readout software determines if an individual black bar value is above or below the red line and interprets it accordingly to a binary number "1" or "0".
Every transmitted message is initialized by the bit sequence "000010" to normalize the length of a single bit, before the start sequence is removed and the first sign begins. The end of the transmission is defined by the sequence "110000". After conversion into letters, the program reveals the correctly transmitted message, as presented by the binary numbers and according letters above the data in Fig.~\ref{figure3}.\\

\setlength{\parindent}{0cm}

{\bf \large DISCUSSION} \\
The biprism interferometer is comparable to the famous double slit experiment in quantum physics \cite{Jonsson1974}. It is well known that interference effects vanish as soon as the separated particle waves are measured before getting superposed \cite{Sonnentag2007,Sanz2005,Hornberger2003,Hackermuller2004}. For that simple reason, our transmission scheme intrinsically includes a high level of security on a quantum level against passive eavesdropping. Here, we would like to discuss in more detail different attacks and why we are the opinion that our quantum wave modulation scheme has a significantly improved security compared to classical transmission. 
\setlength{\parindent}{5mm}

The experiment has two regions with different levels of security. The high security communication distance is between the Wien filter and the coherent partial beam overlap within or close to the entrance of the magnifying quadrupole. There, the coherent partial waves have not overlapped yet, direct measurement by an eavesdropper will only yield two independent beam spots of electrons and the modulated common information is lost. The exact distance depends on the amount of fringes needed for analysis which is related to the width of the superposition, the superposition angle, the electron wavelength and the resolution or magnification in the detection process. In this experiment it is around \SI{38}{mm} which is the distance between the Wien filter's center and the magnifying quadrupole's center. 

The second region with low security is between the beam superposition in the quadrupole and the detector. This distance is \SI{102}{mm}. It can be argued that the partial waves are already superposed and the fringe pattern with the encoded information is visible at any place in different magnifications.\\

\begin{figure}[t]
\centering
\includegraphics[width=0.45\textwidth]{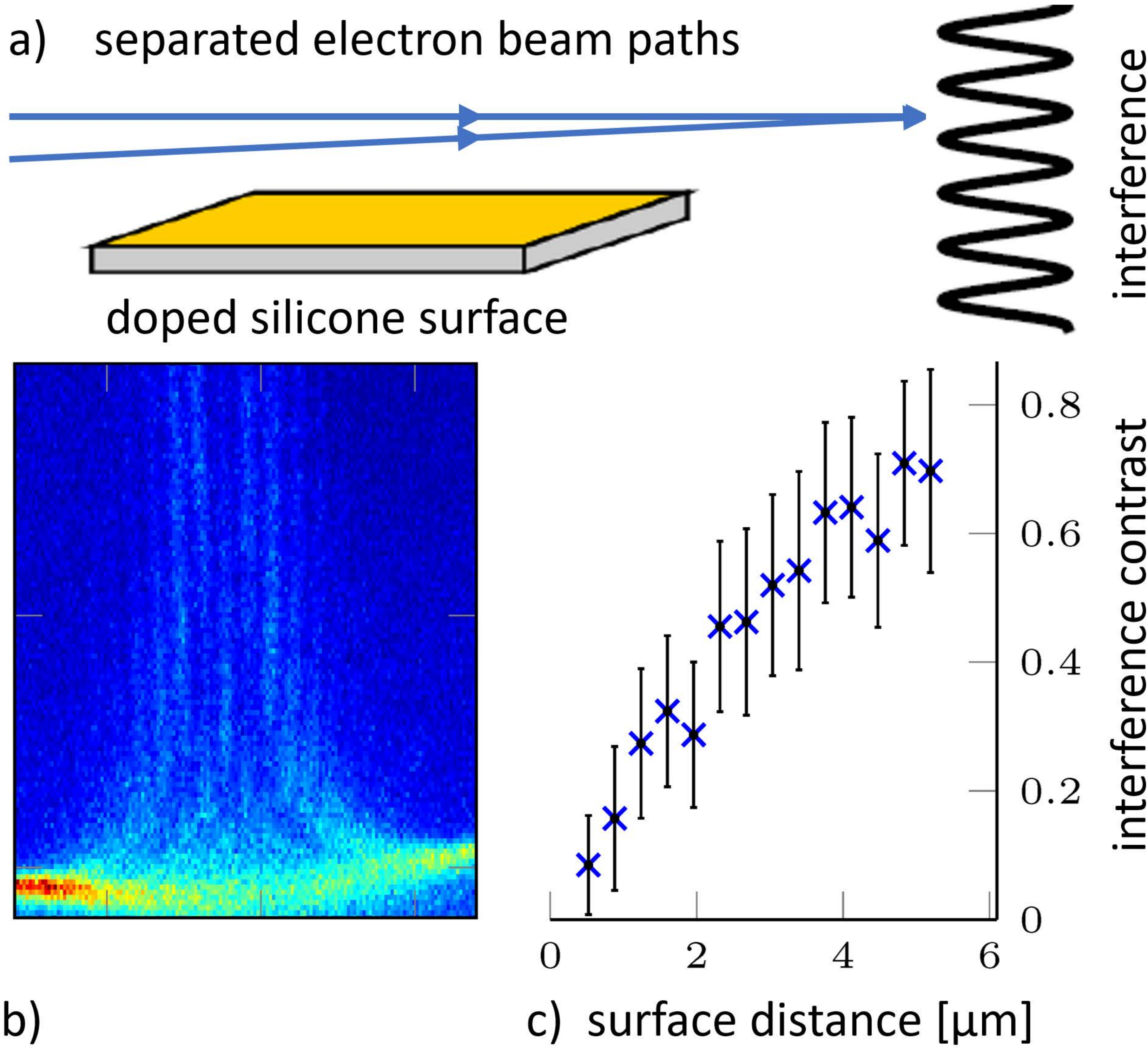} \caption{The effect of an eavesdropper interfering with the communication. a) A semiconducting doped silicon surface is introduced perpendicular to the biprism fibre direction at the lower image edge into the beam path before superposition. b)~Determined fringe pattern on the detector. The bright region indicates the surface. The proximity of such a conducting device decreases the interference contrast due to Coulomb-induced decoherence \cite{Sonnentag2007,Kerker2020,Beierle2018,Scheel2012}. c)~The fringe contrast is plotted versus the vertical distance between the surface and the electrons in the superposition state.}
\label{figure4}
\end{figure}

\noindent {\bf Security based on quantum decoherence:} \\
An eavesdropper in the first (secure) region of the transmission coming close to the electrons in the quantum superposition state and tapping with any kind of measuring device, will cause decoherence due to Coulomb-interaction \cite{Sonnentag2007,Kerker2020}. This reduces or destroys the interference contrast on which our binary coding technique is based on. It is thereby not important if information about the electron waves is actually measured by an external observer or could only in principle be gained from the environment. 
For that reason, any conducting surface can represent the effect of an eavesdropper. The surface atoms (as being the environment) can resolve the which-path information of the electrons if the surface is close enough to the separated coherent beam paths. We studied and compared this Coulomb-induced decoherence with current theoretical models in detail elsewhere \cite{Kerker2020}. It was also experimentally analyzed in \cite{Sonnentag2007,Beierle2018} and theoretically discussed in \cite{Scheel2012}. Here, we demonstrate such an eavesdropping attempt by applying a one-centimeter-long doped silicon surface parallel and below the separated beam lines before superposition, as illustrated in Fig.~\ref{figure4}~a). The resulting fringe pattern on the detector from the aloof electron waves with a beam path separation of $\Delta x =$ \SI{5.0}{\micro\metre} is shown in Fig.~\ref{figure4}~b). Fig.~\ref{figure4}~c) plots the determined interference contrast as a function of the electron distance normal to the semiconducting surface. It can be clearly observed that the contrast drops from about 70\% to zero within \SI{5}{\micro\metre}. This represents the distinct effect of a passive eavesdropper who would be detected immediately and end the communication. The process depends on the beam path separation, surface conductivity and temperature \cite{Sonnentag2007,Kerker2020,Beierle2018}.\\

\begin{figure}[t]
\centering
\includegraphics[width=0.35\textwidth]{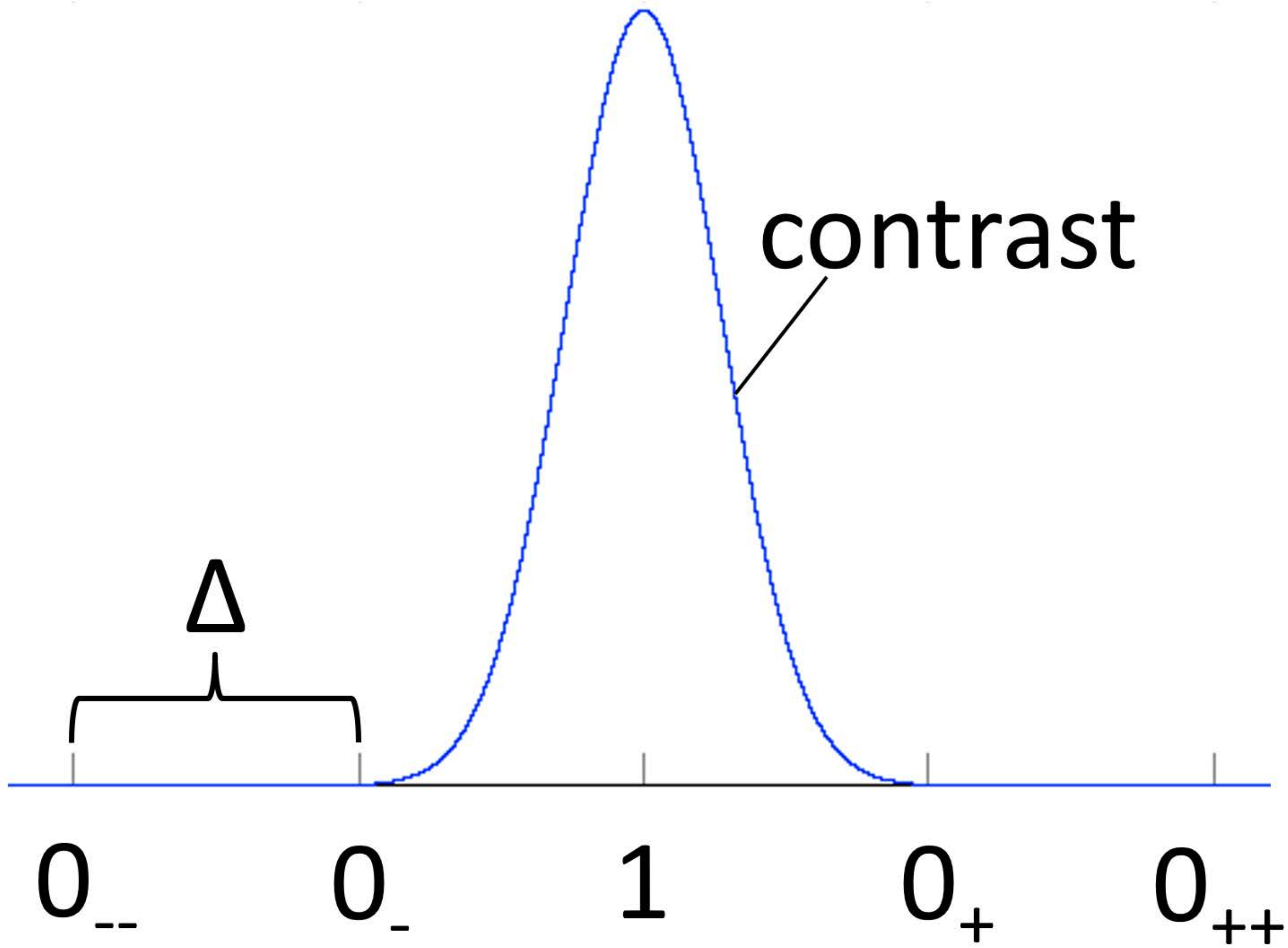} \caption{Schematic illustration of the interference contrast vs.~Wien filter deflector voltage from Fig.~\ref{figure2} to indicate the signal encoding points in the matterwave key distribution protocol. The shift between the separated wave packets from max.~to min.~contrast is $\Delta$. The positions $0_{--}$, $0_-$, $0_+$ and $0_{++}$ indicate the points at multiples of $\Delta$ in both directions.}
\label{figure5}
\end{figure}

\noindent {\bf Matterwave key distribution protocol:} \\
More complicated is to uncover active eavesdropping in a "man in the middle" attack. In this scenario, the interceptor deflects the separated coherent beams before overlap and superposes them on his own detector to measure the high and low contrast interference pattern according to the transmitted signal. Then, he or she applies a second matterwave interferometer and induces an according interference pattern into the optical axis of the receivers magnifying quadrupole and detector. With such a tapping setup, the receiver may not determine the presence of the eavesdropper. However, we present a transmission scheme using the quantum nature of the electron wavepackets to reveal such an attack and accomplish secure data transfer. Our method has similarities to the BB84 protocol for quantum key distributions with photons \cite{Bennett1984}. It utilizes the symmetry of the curve in Fig.~\ref{figure2}, where the contrast value at the indicated point C is equivalent to the one at point A. It leads to two possibilities how a '0' can be encoded, which an eavesdropper cannot distinguish. Let's assume for this scheme that point A and C in Fig.~\ref{figure2} are set further away from the center of the Gaussian curve such as illustrated by the points $0_-$ and $0_+$ in Fig.~\ref{figure5}. They have the same zero contrast value within the error as any other point in the background far away from the wave packets overlap. We also assume an equal negative and positive Wien filter shift $\Delta$ between minimal contrast at point $0_-$ or $0_+$ and the maximum contrast at perfect wave packets overlap (point B in Fig.~\ref{figure2} or point 1 in Fig.~\ref{figure5}). The positions where the partial wave packets are shifted by $\pm 2 \Delta$ from max.~overlap are labeled as $0_{--}$ and $0_{++}$.

For our matterwave key distribution scheme, let's assume the implementation of a second Wien filter for signal encoding at the sender ($W_s$) and a further one at the receiver's end ($W_r$), directly before the magnifying quadrupole. $W_s$ encodes the message by shifting the wave packets by $\pm \Delta$ or leaving it as it is. The receiver's $W_r$ randomly does the same ($\pm \Delta$ or no shift). After several turns, the sender and receiver exchange via a public channel for each bit how they have shifted (but not if they sent or measured a '0' or a '1'). They use only those bits as an encryption key, where both did no shift, or where they did the opposite shift. This way, the receiver reverses the shift of the sender and measures the originally transmitted bit. For completeness it is worth to mention that in principle the first and second Wien filter of the sender can be combined to one Wien filter.

As a next step, the sender and receiver publicly exchange the remaining sent/received bits (or a separate small subset of the data) together with the applied shifts. Comparing the results are used to reveal a possible eavesdropper. To provide an example, if the sender sends a "1" (max.~contrast) and shifts it with $W_s$ by $-\Delta$ to a '0' bit at point $0_-$ (see Fig.~\ref{figure5}) and the receiver chooses to shift with $W_r$ also by $-\Delta$, he will end up at point $0_{--}$ and measure a '0' bit (min.~contrast). If an eavesdropper is in line and chooses 'no shift' in the Wien filter, he or she measures a '0' not knowing if it was sent as a $0_{--}$, $0_-$, $0_+$ or $0_{++}$. In case he chooses to send it as a $0_+$ the receiver (after subtracting $\Delta$) will end up at the max.~of the contrast curve at point '1' and measure a '1' bit. Since this is the opposite value as without the eavesdropper this interception can be revealed after the public data comparison. All possible setting on the sender and receiver site are summarized in Table~\ref{table1}. 

It also demonstrates the large uncertainty an eavesdropper has. If Table~\ref{table1} is considered a 3$\times$3 matrix with the elements (row nr., column nr.) the three cases for the eavesdropper (E) can be analyzed in more detail:\\
Case 1, (E) has set his Wien filter on "no shift" (no): \\
(E) measures a 0: it could be (11), (12), (21), (23), (32), (33).
(E) measures a 1: it could be (13), (22), (31). \\
Case 2, (E) has set his Wien filter on  $-\Delta$:\\
(E) measures a 0: it could be (11), (12), (13), (21), (22), (31), (33).
(E) measures a 1: it could be (23), (32). \\
Case 3, (E) has set his Wien filter on  $+\Delta$:\\ 
(E) measures a 0: it could be (11), (13), (22), (23), (31), (32), (33).
(E) measures a 1: it could be (12),(21). 

For that reason, even in the best case when (E) measures a '1', he or she has only a 50-50 chance to correctly forward the bit to the receiver. Without the information about the applied shifts from the sender, the eavesdropper cannot decode the transmitted sequence.
\begin{center}
\begin{table}[t]
\centering
\begin{tabular}{|c||c|c|c|}
\hline
 & $0_-$ & $1$ & $0_+$\\
\hline
\hline
$-\Delta$ & $0_{--}\rightarrow $\begin{small}\begin{tabular}{c|c} $-\Delta$ & $0_{---}$\\ \hline $no$ & $0_{--}$ \\ \hline $+\Delta$ & $0_-$ \end{tabular}\end{small} & $0_-\rightarrow $\begin{small}\begin{tabular}{c|c} $-\Delta$ & $0_{--}$\\ \hline $no$ & $0_{-}$ \\ \hline $+\Delta$ & $1$ \end{tabular}\end{small} & $1\rightarrow $\begin{small}\begin{tabular}{c|c} $-\Delta$ & $0_{-}$\\ \hline $no$ & $1$ \\ \hline $+\Delta$ & $0_+$ \end{tabular}\end{small}\\
\hline
$no$ & $0_{-}\rightarrow $\begin{small}\begin{tabular}{c|c} $-\Delta$ & $0_{--}$\\ \hline $no$ & $0_{-}$ \\ \hline $+\Delta$ & $1$ \end{tabular}\end{small} & $1\rightarrow $\begin{small}\begin{tabular}{c|c} $-\Delta$ & $0_{-}$\\ \hline $no$ & $1$ \\ \hline $+\Delta$ & $0_+$ \end{tabular}\end{small} & $0_+\rightarrow $\begin{small}\begin{tabular}{c|c} $-\Delta$ & $1$\\ \hline $no$ & $0_{+}$ \\ \hline $+\Delta$ & $0_{++}$ \end{tabular}\end{small}\\
\hline
$+\Delta$ & $1\rightarrow $\begin{small}\begin{tabular}{c|c} $-\Delta$ & $0_{-}$\\ \hline $no$ & $1$ \\ \hline $+\Delta$ & $0_+$ \end{tabular}\end{small} & $0_+\rightarrow $\begin{small}\begin{tabular}{c|c} $-\Delta$ & $1$\\ \hline $no$ & $0_{+}$ \\ \hline $+\Delta$ & $0_{++}$ \end{tabular}\end{small} & $0_{++}\rightarrow $\begin{small}\begin{tabular}{c|c} $-\Delta$ & $0_{+}$\\ \hline $no$ & $0_{++}$ \\ \hline $+\Delta$ & $0_{+++}$ \end{tabular}\end{small}\\
\hline 
\end{tabular} 
\caption{Possible combinations of wave packet shifts from the sender and receiver for the matterwave key distribution protocol and the corresponding bit measurement. The columns are labeled by the points $0_-$, 1 or  $0_+$ (see Fig.~\ref{figure5}) where the sender sends the original bit '0' or '1' with the first Wien filter. The rows provide the possible shifts ($-\Delta$, 'no' (no shift) and $+\Delta$) for signal encoding with the sender's second Wien Filter $W_s$. The first value in the boxes provides the signal state after the encoding by the sender ($0_{--}$ to $0_{++}$) and behind the arrow the receivers states after the decoding shifts by $W_r$ of $-\Delta$, 'no' and $+\Delta$.}
\label{table1}
\end{table}
\end{center} 

\vspace{-8mm}
One final loophole that may be discussed is when the transmission is performed with significantly more signal than required for a contrast determination, (E) could manage to completely capture all signal in an active attack and also to coherently split the beam multiple times. This would allow to connect several interferometers that are set on all possible wave packet shifts ($-\Delta$, 'no', $+\Delta$). After the public exchange of the sender and receivers shift setting, (E) could decode the message. However, (E) still had to decide which state to forward to the receiver during transmission and will do this wrongly in several cases. For that reason, also in this case our key distribution protocol would reveal the presence of the interceptor. 

It needs to be mentioned that there is still an important difference between our matterwave scheme and quantum protocols with photons such as BB84 \cite{Bennett1984}. The BB84 protocol is based on the uncertainty in the measurement of the single photon quantum state. Our scheme depends on the quantum nature and symmetry of a matterwave packet and therefore is based on the single particle interference visible in a multiparticle fringe pattern. However, our security argument is still based on a single-particle quantum effect. The message is encoded in the interference that relies on the wave nature of each electron interfering with itself. It forms a full interference pattern as a probability distribution that is dependent on the Wien filter settings (binary number “0” or “1”). As discussed and demonstrated, an eavesdropper introduces decoherence that in turn destroys the wave features of each electron individually. They cannot form a wave-packet anymore and our whole information transfer scheme and key distribution protocol breaks down. The more the interceptor interferes to get information the more information is lost due to his disturbance of the quantum system. This in turn reduces the contrast of the individual single electron quantum probability distribution. As a consequence, the interference pattern and the communication are interrupted. In the final readout process we need to sum over an ensemble of particles to reveal this distribution. Here lies also a major difference in our technique towards single photon protocols, and maybe also a small drawback. To reveal the information (the interference contrast) encoded in every electron's probability distribution, the measurement of a certain amount of particles is required. For that reason, the electron count rate of the source needs to be calibrated and known by the receiver to a avoid a coherent and undetected beam splitting by an eavesdropper.

The single particle character can further be pointed out in a thought experiment where already a single electron in an idealized interferometer with 100\% contrast and strong fringe magnification can (with a certain probability) detect an eavesdropper. The probability distribution in such an interferometer would predict a few clear dark and bright fringes. If our setup is set to send a "1"-bit (full contrast), and there is no eavesdropper, the electron will never be measured in the area on the detector around the centers of the dark fringes minima. In case the electron is detected there, an eavesdropper tapped the line and reduced the contrast.

In summary, we give experimental evidence that the fundamental quantum principles of the electron wave nature and decoherence significantly improve the transmission security. Since our Wien filter scheme does not shift the fringe position on the detector and the electric and magnetic fields cancel each other, the coherent separated electron pathways are not altered in any classical way. For that reason, also the electromagnetic fields from the electron beams is classically not varied when switching between "0" and "1" making it impossible to read the information with any classical measuring device while the partial beams are still separated. The only change is the group velocity of the wave packets that classically changes the arrival times of the particles. However, since the electron emission is a Poisson distributed statistical process \cite{Schuetz2014}, the individual electron starting time is unknown, making the electron time differences not accessible. The security aspect of our scheme is therefore also a direct consequence of the random quantum tunneling process through the field emitters Coulomb barrier in the Schottky effect \cite{Orloff2008}. We show that passive eavesdropping is prohibited due to the introduction of decoherence \cite{Kerker2020,Beierle2018,Scheel2012} and active eavesdropping can be prevented by a key distribution scheme based on the wave packets symmetry. Further analysis of the influence on the transmission security of technical parameters such as interference contrast, count rate fluctuations, beam separation, noise or electron energy spread are beyond this proof-of-principle work and will be the aim of upcoming studies.\\

\noindent {\bf Future prospects and technical applications:} \\
Our proof-of-principle experiment has yet a secure distance which is probably too short for direct technical applications. It certainly cannot compete with the established state-of-the-art photon quantum communication schemes. They allow secure intra-country distances accessible with glass fibers \cite{Stucki2009} or even over long, intercontinental distances via satellite communication \cite{Liao2017}. However, the save signal transmission distance in our scheme could be extended significantly by increasing the beam path separation. Coherent electron beam splitting up to \SI{300}{\micro\meter} have been reported \cite{Schmid1985}. With the same superposition angle, secure transmission of \SI{3.8}{m} is feasible with such a separation with state-of-the-art technology. It can be further extended to $\sim$ \SI{6}{m} by positioning the Wien filter closer to the tip, even between the tip and the biprism \cite{Nicklaus1993}. Also, the data transfer rate has the potential to be improved significantly. We have been conservative with our signal rate, to preserve our source. However, driving our system to the maximum could increase the transmission rate by at least an order of magnitude. Optimizing the whole setup and the electron beam sources \cite{Schuetz2014} can potentially gain another factor 100. 

Potential applications of our technique are short range communication in environments, where photons cannot be applied due to an intensive light background on the sensors. It is also conceivable that in a low noise environment such as in space significantly larger beam paths separations and transmission distances are realistic. 

The quest for novel quantum techniques for communication is important in various fields of modern science and technology. Here, we demonstrated in a proof-of-principle experiment that it is possible to perform a true quantum modulation on a matterwave for signal transfer in a biprism interferometer. The information is transmitted by a longitudinal wave packet shift introduced by a Wien filter. We show that decoherence plays an important role for the security aspect and present a matterwave key distribution protocol to prevent a direct eavesdropping attack. It is also emphasized that the Wien filter has no light wave optical analog. It has an electron optical refractive index of one \cite{Hasselbach2010} and is able to shift the wave packets longitudinally without shifting the transverse phase of the fringe pattern in contrast to light optics. This is a result of the compensating phase shifts from the magnetic and electric Aharonov-Bohm effects \cite{Hasselbach2010} for charged particles. For that reason, our scheme cannot be performed with laser light nor with neutral atoms and presents a unique method for electron waves. It is a separate class of information transfer technique based on matterwave quantum features and indicates significant improvement in safety compared to classical communication.\\
\setlength{\parindent}{0cm}

{\bf \large METHODS} 

We apply the Wien filter to shift the wave packets of the separated beam paths as described in detail in \cite{Pooch2018,Nicklaus1993}. The Wien filter introduces a longitudinal shift of $\Delta y = \frac{L \, \Delta x \, U_{WF}}{2 \, D \, U_{tip}} \label{eq1}$, where $L$ is the length of the Wien filter deflector plates, $\pm U_{WF}$ the applied voltage, $D$ is the distance between them and $U_{tip}$ the electron acceleration voltage \cite{Pooch2018}. The longitudinal coherence length is the width of the wave packets and can be calculated with $l_c = \frac{2 \, U_{tip} \, \lambda}{\pi \, \Delta E}$, with $\lambda $ being the electron de Broglie wavelength at $U_{tip}$ \cite{Pooch2018}.
\setlength{\parindent}{5mm}

The following parameters were applied for the signal transmission in Fig.~\ref{figure3}: $L=$ \SI{11.75}{mm}, $\Delta x = $ \SI[separate-uncertainty = true]{2.9 \pm 0.4}{\micro\metre} at a biprism voltage of \SI{-0.104}{V} and a combining quadrupole voltage of \SI{-15.9}{V}, $D=$ \SI{8.75}{mm}, $U_{tip}=$ \SI{1000}{V}, $\lambda =$ \SI{38.8}{pm}, $\Delta E =$ \SI[separate-uncertainty = true]{377(40)}{meV}, $l_c =$ \SI[separate-uncertainty = true]{66(7)}{nm}. The distances from the tip to the biprism, combining quadrupole (center), Wien filter (center) and magnifying quadrupole (center) have been $d_{tip-bp}=$ \SI{83}{mm}, $d_{tip-QP1}=$ \SI{125}{mm}, $d_{tip-WF}=$ \SI{237}{mm} and $d_{tip-QPmag}=$ \SI{275}{mm}, respectively. For state 2 (high contrast) and state 1 (low contrast) the Wien filter deflector voltages $U_{WF \, high}=$ \SI{-15}{V} and $U_{WF \, low}=$ \SI{-45}{V} were set. The wavepackets are shifted by $\Delta y = $ \SI{58}{nm} at each state change. The corresponding currents in the Wien filter coil are set in a way that the beam deflection vanishes and the Wien filter is in the matched state. The whole setup is in an ultrahigh vacuum at a pressure of \SI{1e-10}{mbar} and shielded by a mu-metal tube. 

For every deflector voltage in the Wien curve of Fig.~\ref{figure2}, five pictures with 250.000 particle events each were taken and averaged. Every single measurement lasts only approximately one minute to avoid long-term drifts. The contrast value data points in Fig.~\ref{figure2} and Fig.~\ref{figure3} were determined by summing up the signal along the fringe direction to form a histogram. The fit function \mbox{$I(x)=I_0\cdot\left(1+C\cdot\cos(\frac{2\pi x}{s}+\phi_0)\right)\cdot {\rm sinc}^2(\frac{2\pi x}{s_1}+\phi_1)$} was then applied as described in more detail elsewhere \cite{Pooch2017,Pooch2018}. Here, $C$ describes the interference contrast and $s$ the fringe distance. Further fitting parameters are the phases $\phi_0$, $\phi_1$, the average intensity $I_0$ and the width of the interference pattern $s_1$. 

For the contrast analysis in Fig.~\ref{figure4} c), the interference was determined as a function of the surface distance. Each contrast point is evaluated by slicing a horizontal rectangular section of the image normal to the fringe orientation with a height of \SI{400}{\nano\meter}. The counts in each slice were summed up to form a histogram that was fitted with the above function. The contrast was normalized to the undisturbed value at distances higher than \SI{20}{\micro\m} to the semiconducting surface.

\setlength{\parindent}{5mm}

\vspace{6mm}
\setlength{\parindent}{0cm}
{\bf \large ACKNOWLEDGEMENTS} \\
This work was supported by the Vector Stiftung through the initiative "MINT-Innovationen". Work at the Molecular Foundry was supported by the Office of Science, Office of Basic Energy Sciences, of the U.S. Department of Energy under Contract No.~DE-AC02-05CH11231. We also acknowledge support by the Deutsche Forschungsgemeinschaft through the research grant STI 615/3-1.\\

{\bf \large AUTHOR CONTRIBUTIONS}\\
A.S.~conceived the research and wrote the manuscript. R.R.~and N.K.~conducted the measurements and performed the data evaluation.\\

{\bf \large COMPETING INTERESTS}\\
The authors declare no competing interests.

\end{document}